\documentclass[12pt]{article}
\usepackage{amsmath}
\usepackage{amsthm}
\usepackage{amssymb}

\newtheorem{lemma}{Lemma}
\newtheorem{thm}{Theorem}
\theoremstyle{remark}
\newtheorem*{remark}{Remark}
\newcommand{\abs}[1]{|#1|}
\newcommand{\norm}[1]{\| #1\|}
\newcommand{\alg}[1]{\mathfrak{#1}}
\newcommand{\hil}[1]{\mathcal{#1}}
\newcommand{\weyl}{\mathfrak{A}[\mathbb{R}^{2}]}
  \title{The Einstein-Podolsky-Rosen State Maximally Violates Bell's
    Inequalities} 
\author{Hans Halvorson \\
{\small Department of Philosophy, University of Pittsburgh } \\ 
{\small e-mail: hphst1@pitt.edu } }
\date{}
\begin{document}
\maketitle
\vspace{-1.5em} 
\begin{abstract} In their well-known argument against the completeness of quantum
  theory, Einstein, Podolsky, and Rosen (EPR) made use of a state that
  strictly correlates the positions and momenta of two particles.  We
  prove the existence and uniqueness of the EPR state as a normalized,
  positive linear functional of the Weyl algebra for two degrees of
  freedom.  We then show that the EPR state maximally violates Bell's
  inequalities. \end{abstract}

\section{Introduction}
In their well-known argument against the completeness of quantum
theory, Einstein, Podolsky, and Rosen (EPR) make use of a state that
strictly correlates the positions and momenta of two
particles~\cite{epr}.  However, since these strict correlations
require the EPR state to be a common ``eigenstate'' of a pair of
continuous spectrum observables, viz., $\hat{x}_{1}-\hat{x}_{2}$ and
$\hat{p}_{1}+\hat{p}_{2}$, the problematic mathematical status of the
EPR state has caused difficulties for two recent attempts to assess
its foundational significance.

First, it has so far remained an open question whether the EPR state
has nonlocal features similar to those of Bohm's spin-$1/2$ singlet
state.  On the one hand, J. S.  Bell~\cite{bell} has shown that the
correlations in the EPR state admit a local hidden variable model, at
least for certain simple functions of the positions and momenta of the
two particles.  On the other hand, Cohen~\cite{cohen} has claimed to
show that the EPR state \emph{does} predict nonlocal correlations
between more general observables associated with the respective
particles.  Unfortunately, Cohen's argument is unsatisfactory, since
it attempts to use the fact that any entangled vector state violates
Bell's inequalities to draw a conclusion about the EPR state---which
is not a vector state in the standard Schr{\"o}dinger representation.

Second, Arens and Varadarajan~\cite{vara} have recently given a
characterization of states with EPR-type correlations.  Ironically,
however, their characterization does not extend to correlations
between observables with continuous spectra; in particular, their
characterization does not extend to the original EPR state.

In this letter, we resolve \emph{both} difficulties noted above.
First, we show that the EPR state has a mathematically unproblematic
definition as a normalized, positive linear functional on the Weyl
algebra for two degrees of freedom.  Second, we show that the EPR
state maximally violates Bell's inequalities.  Finally, we show that
the original EPR state satisfies the characterization of EPR-type
states recently given by Werner~\cite{werner}, which naturally extends
the characterization given by Arens and Varadarajan.

\section{Existence and uniqueness of the EPR state}
Let $\alg{A}[\mathbb{R}^{2}]$ denote the Weyl algebra over
$\mathbb{R}^{2}$.  That is, $\weyl$ is the $C^{*}$-algebra generated
by unitary operators $W(\mathbf{a})$, $\mathbf{a}\in \mathbb{R}^{2}$,
satisfying the Weyl relations
\begin{equation} W(\mathbf{a})W(\mathbf{b})=\exp \{ i\sigma
  (\mathbf{a},\mathbf{b})\} W(\mathbf{a}+\mathbf{b})
  , \label{weyl} \end{equation} with \begin{equation} 
\sigma ((a,b),(a',b'))={\textstyle \frac{1}{2} }(ab'-ba') , 
\qquad a,b \in \mathbb{R}.\end{equation}
(See~\cite{manuceau,petz} for more
details.)  Let $\alg{A}[\mathbb{R}^{4}]$ denote the Weyl algebra over
  $\mathbb{R}^{4}$.  (In this case, the Weyl generators satisfy the
  analogue of Eq.~\ref{weyl} with $\sigma \oplus \sigma$ replacing $\sigma$.)  Since
$\alg{A}[\mathbb{R}^{2n}]$ is simple $(n=1,2)$, all representations of
$\alg{A}[\mathbb{R}^{2n}]$ are faithful.  We say that a linear
functional $\omega$ of a $C^{*}$-algebra $\alg{A}$
is a \emph{state} just in case $\omega$ is positive and normalized.  
 
The observables for a composite system consisting of a pair of
one-dimensional particles are given by the self-adjoint elements of
the spatial tensor product $\weyl \otimes \weyl$.  If $\pi$ is a
regular representation of $\weyl \otimes \weyl$, then Stone's theorem
entails the existence of self-adjoint operators
$\hat{x}_{i},\hat{p}_{i}\, (i=1,2)$ such that
\begin{eqnarray}
\pi (W(a,0)\otimes I)=e^{ia\hat{x}_{1}}, \qquad \pi (W(0,b)\otimes
I)=e^{ib\hat{p}_{1}}, \\
\pi (I\otimes W(c,0))=e^{ic\hat{x}_{2}}, \qquad  \pi (I\otimes
W(0,d))=e^{id\hat{p}_{2}}. \end{eqnarray}  
Heuristically, the EPR state assigns a dispersion-free value $\lambda$
to the relative position $\hat{x}_{1}-\hat{x}_{2}$ of the two
particles, and a dispersion-free value $\mu$ to the total momentum
$\hat{p}_{1}+\hat{p}_{2}$ of the two particles.  (EPR themselves
chose the value $\mu =0$.)  Accordingly, the EPR
state assigns the value $e^{ia\lambda}$ to
$e^{ia(\hat{x}_{1}-\hat{x}_{2})}=e^{ia\hat{x}_{1}}e^{-ia\hat{x}_{2}}$
and the value $e^{ib\mu}$ to
$e^{ib(\hat{p}_{1}+\hat{p}_{2})}=e^{ib\hat{p}_{1}}e^{ib\hat{p}_{2}}$.
We show now that these latter conditions suffice to define a unique
EPR state of $\weyl \otimes \weyl$.

\begin{thm} For each pair of real numbers $\lambda ,\mu $, there is a unique state
  $\omega$ of $\weyl \otimes \weyl$ such that
  \begin{equation}
\omega (W(a,0)\otimes W(-a,0))=e^{ia\lambda}  ,\qquad 
\omega (W(0,b)\otimes W(0,b))=e^{ib\mu}  ,\end{equation}  
for all $a,b\in \mathbb{R}$.  Moreover, $\omega$ is pure.  \end{thm}

\begin{proof} (Existence) We presuppose the canonical
  isomorphism between $\weyl \otimes \weyl$ and
  $\alg{A}[\mathbb{R}^{4}]$~(see \cite[3.4.1]{manuceau}).  Every state
  $\omega$ on $\alg{A}[\mathbb{R}^{4}]$ gives rise to a function
  $G:\mathbb{R}^{4}\mapsto \mathbb{C}$ via the equation
  $G(\mathbf{x})=\omega (W(\mathbf{x}))$.  Conversely, a function
  $G:\mathbb{R}^{4}\mapsto \mathbb{C}$ gives rise to a state $\omega$
  on $\alg{A}[\mathbb{R}^{4}]$ just in case $G(0)=1$ and the map
  $\mathbf{x},\mathbf{y}\mapsto G(\mathbf{x}-\mathbf{y}) \exp \{
  -i(\sigma \oplus \sigma ) (\mathbf{x},\mathbf{y})\}$ is a positive
  definite kernel~\cite[p.~18]{petz}.

Define $G$ by 
\begin{equation}   G(a,b,c,d) = \delta (a+c)\delta (b-d) \, 
e ^{ia\lambda}e^{ib\mu} , \end{equation}
where $\delta$ is the characteristic function of $\{ 0\}$.  Obviously,
$G(0)=1$.  Now define
$F:\mathbb{R}^{4}\times \mathbb{R}^{4} \mapsto \mathbb{C}$
by \begin{equation}
F(\mathbf{x},\mathbf{y})=G(\mathbf{x}-\mathbf{y}) \exp \{ -i(\sigma \oplus \sigma ) 
(\mathbf{x},\mathbf{y}) \} . \end{equation}  To see that $F$ is
positive-definite, let $\{ z_{1},\dots ,z_{n} \}
\subseteq \mathbb{C}\backslash \{ 0\}$ and let $\{ \mathbf{x}_{1},\dots
,\mathbf{x}_{n}\} \subseteq \mathbb{R}^{4}$ with
$\mathbf{x}_{j}=(a_{j},b_{j},c_{j},d_{j})$.
Fix $j,k$.  It follows then that \begin{equation}  
F(\mathbf{x}_{j},\mathbf{x}_{k})=
\delta (a_{j}-a_{k}+c_{j}-c_{k})\,\delta
  (b_{j}-b_{k}+d_{k}-d_{j})\, \alpha _{j} \overline{\alpha}_{k}
  , \label{inspector} \end{equation}
where \begin{equation}
\alpha _{j}=\exp \{ i(a_{j}\lambda +b_{j}\mu ) \} \exp \{ 
i[b_{j}(a_{j}+c_{j})-a_{j}(b_{j}-d_{j})] \} . \label{gadget} \end{equation}  Define a
relation $R$ on $\{ 1,\dots ,n\}$ by
\[ (j,k)\in R \quad \mbox{iff.} \quad F(\mathbf{x}_{j},\mathbf{x}_{k})
\neq 0 .\] An inspection of Eq.~\ref{inspector} shows that $R$ is an
equivalence relation.  Thus, there are disjoint subsets $S_{1},\dots
,S_{m}$ of $\{ 1,\dots ,n\}$ such that $R=\cup _{i=1}^{m}(S_{i}\times
S_{i})$, and
\begin{eqnarray}
\lefteqn{ \sum _{j=1}^{n}\sum
  _{k=1}^{n}z_{j}\overline{z}_{k}F(\mathbf{x}_{j},\mathbf{x}_{k}) \:= \:
\sum _{(j,k)\in R} z_{j}\overline{z}_{k}\alpha _{j}\overline{\alpha}_{k} } \qquad \\
&=& \sum _{(j,k)\in S_{1}\times S_{1}}z_{j}\alpha
  _{j}\overline{z_{k}\alpha _{k}}
\: + \: \cdots \: + \: \sum _{(j,k)\in S_{m}\times S_{m}}z_{j}\alpha _{j}\overline{z_{k}\alpha 
_{k}} \\
&=& \Bigl| \sum _{j\in S_{1}}z_{j}\alpha _{j} \Bigr| ^{2} + \cdots + \Bigl|
\sum _{j\in S_{m}}z_{j}\alpha _{j} \Bigr| ^{2} \: \geq \: 0 .\end{eqnarray}
Therefore, $F$ is positive-definite.   

(Uniqueness) Let $\alg{A}=\weyl \otimes \weyl$ and let $\rho$ be a
state of $\alg{A}$ such that
\begin{equation}
  \rho (W(a,0)\otimes W(-a,0))=e^{ia\lambda } ,\qquad 
\rho (W(0,b)\otimes W(0,b))=e^{ib\mu } , \label{defn1} \end{equation} 
for all $a,b \in \mathbb{R}$.  Fix $s,t\in \mathbb{R}$, let $A=W(s,0)\otimes
W(-s,0)$, and let $B=W(0,t)\otimes W(0,t)$.  Thus, $\abs{\rho (A)}=1=\norm{A}$ and $\abs{\rho
    (B)}=1=\norm{B}$.  Since $A,B$ are unitary, it follows
from~\cite[p.~305]{ande} that \begin{eqnarray} 
  \rho (AX) &=&\rho (XA)\:=\: \rho (X)\rho (A) , \label{add1} \\
\rho (BX) &=&\rho (XB) \: =\: \rho (X) \rho (B) ,\label{add2} \end{eqnarray} for
any $X\in \alg{A}$.  Let $W(a,b)\otimes W(c,d)\in \alg{A}$.  Then, using the Weyl
relations, we have \begin{eqnarray} \lefteqn{ [W(s,0)\otimes
    W(-s,0)][(W(a,b)\otimes
    W(c,d)] =} \qquad \nonumber \\
  && e^{isb}e^{-isd}[W(a,b)\otimes W(c,d)][W(s,0)\otimes W(-s,0)]
  \label{first} ,\end{eqnarray} and \begin{eqnarray}
  \lefteqn{ [W(0,t)\otimes W(0,t)][W(a,b)\otimes W(c,d)] =} \qquad \nonumber \\
  && e^{ita}e^{itc}[W(a,b)\otimes W(c,d)][W(0,t)\otimes W(0,t)]
  \label{second}. \end{eqnarray} Using Eqs.~\ref{defn1},~\ref{add1},
and~\ref{first}, it follows that
  \begin{equation}
\rho (W(a,b)\otimes W(c,d))=e^{isb}e^{-isd}\rho (W(a,b)\otimes W(c,d)) .
\end{equation}
Since this is true for all $s\in \mathbb{R}$, it follows that $\rho
(W(a,b)\otimes W(c,d))=0$ when $d\neq b$.  Similarly,
Eqs.~\ref{defn1},~\ref{add2}, and~\ref{second} entail that
\begin{equation}
\rho (W(a,b)\otimes W(c,d))=e^{ita}e^{itc}\rho (W(a,b)\otimes W(c,d))
.\end{equation}
Since this is true for all $t\in \mathbb{R}$, it follows that 
$\rho (W(a,b)\otimes W(c,d))=0$ when
$c\neq -a$.  When $d=b$ and $c=-a$, we have \begin{eqnarray}
\lefteqn{\rho (W(a,b,c,d)) = \rho (W(a,b,-a,b)) } \qquad \qquad \\
&=& \rho ([W(a,0)\otimes W(-a,0)][W(0,b)\otimes W(0,b)]) \\
&=& \rho (W(a,0)\otimes W(-a,0)) \,\rho (W(0,b)\otimes W(0,b)) \\
&=& e^{ia\lambda }e^{ib\mu} .\end{eqnarray}
Thus, $\rho$ agrees with $\omega$ on all Weyl operators.
Since the values of a state on the Weyl operators fixes its values on
$\alg{A}$, it follows that $\rho =\omega$.  

(Purity) Let $\alg{B}$ denote the abelian subalgebra of $\alg{A}$
generated by $W(s,0)\otimes W(-s,0)$ and $W(0,t)\otimes W(0,t)$, with
$s,t\in \mathbb{R}$.  We have seen that $\omega |_{\alg{B}}$ is
multiplicative, and hence is a pure state.  Thus, $\omega |_{\alg{B}}$
has an extension to a pure state $\rho$ on $\alg{A}$.  On the other
hand, we have shown that $\omega |_{\alg{B}}$ has a unique extension.
Therefore, $\omega =\rho$, and $\omega$ is pure. \end{proof}

\section{Maximal Bell correlation of the EPR state}
We first recall some pertinent definitions concerning Bell correlation
for an arbitrary pair of commuting $C^{*}$-algebras.  Suppose then
that $\alg{A}_{1},\alg{A}_{2}$ are mutually commuting subalgebras of a
$C^{*}$-algebra $\alg{A}$, and let $\omega$ be a state of $\alg{A}$.
We set
\begin{eqnarray*} \hil{T}(\alg{A}_{1},\alg{A}_{2}) &\equiv & \{
  {\textstyle \frac{1}{2} }[A_{1}(B_{1}+B_{2})+A_{2}(B_{1}-B_{2})]: \\
  && \quad A_{i}=A_{i}^{*}\in \alg{A}_{1}, B_{i}=B_{i}^{*}\in
  \alg{A}_{2}, \norm{A_{i}}\leq 1, \norm{B_{i}}\leq 1 \}
  .\end{eqnarray*} Elements of $\hil{T}(\alg{A}_{1},\alg{A}_{2})$ are
called \emph{Bell operators} for $\alg{A}_{1},\alg{A}_{2}$.
Following~\cite[p.~223]{summers}, we say that the maximal Bell
correlation of the pair $\alg{A}_{1},\alg{A}_{2}$ in the state
$\omega$ is
  \begin{equation} \beta (\omega ,\alg{A}_{1},\alg{A}_{2}) \equiv \sup
    \{ \omega (R): R\in \hil{T}(\alg{A}_{1},\alg{A}_{2}) \}
 .\end{equation}
It follows that $\beta (\omega ,\alg{A}_{1},\alg{A}_{2}) \in
 [1,\sqrt{2}]$~\cite[Prop.~5.2]{summers}.  
If $\beta (\omega ,\alg{A}_{1},\alg{A}_{2})>1$, we
 say that $\omega$ violates a Bell
 inequality, or is \emph{Bell correlated} across
 $\alg{A}_{1},\alg{A}_{2}$.  
If $\beta (\omega ,\alg{A}_{1},\alg{A}_{2})=\sqrt{2}$, we say that 
$\omega$ is \emph{maximally Bell correlated} across
 $\alg{A}_{1},\alg{A}_{2}$.  In this context, Bell's theorem~\cite{bellbook}
 is the statement that a local hidden variable model of the
 correlations that $\omega$ dictates between $\alg{A}_{1}$ and
 $\alg{A}_{2}$ is possible only if $\beta (\omega
,\alg{A}_{1},\alg{A}_{2})=1$.  
 
In order to demonstrate that the state $\omega$ is Bell correlated
across $\alg{A}_{1},\alg{A}_{2}$, we may wish to pass to the GNS
representation $(\pi ,\hil{H},\Omega )$ induced by $\omega$ (thereby
giving us access to the tools of the theory of von Neumann algebras).
If $\pi$ is faithful, we may choose observables from the weak closures
$\pi (\alg{A}_{1})'', \pi (\alg{A}_{2})''$ in computing the maximal
Bell correlation.

\begin{lemma} Let $(\pi ,\hil{H},\Omega )$ be the GNS
  representation of $\alg{A}$ induced by $\omega$, and let
  $\alg{R}_{i}=\pi (\alg{A}_{i})''$.  If $\pi$ is faithful, then
  \begin{equation} \beta (\omega ,\alg{A}_{1},\alg{A}_{2})=\sup \{ \,
   \langle \Omega ,R\Omega \rangle : R\in
   \hil{T}(\alg{R}_{1},\alg{R}_{2}) \} .\label{wstar} \end{equation}  
\label{cstar} \end{lemma}

\begin{proof} Since $\pi$ is faithful, we have
  \begin{equation} \beta (\omega ,\alg{A}_{1},\alg{A}_{2})=\sup \{
    \,\langle \Omega ,R\Omega \rangle :R\in \hil{T}(\pi
    (\alg{A}_{1}),\pi (\alg{A}_{2})) \} .\end{equation} Using the
  Kaplansky density theorem, and the fact that multiplication is
  jointly continuous (in the strong operator topology) on bounded
  sets, it follows that $\hil{T}(\alg{R}_{1},\alg{R}_{2})$ is
  contained in the strong-operator closure of $\hil{T}(\pi
  (\alg{A}_{1}),\pi (\alg{A}_{2}))$.  The conclusion then follows
  immediately.  \end{proof}

In the case of present interest, we take $\alg{A}_{1}\equiv \weyl
\otimes I$ and $\alg{A}_{2}\equiv I\otimes \weyl$, so that
$\alg{A}_{1},\alg{A}_{2}$ are mutually commuting subalgebras of
$\alg{A}\equiv \weyl \otimes \weyl$.  We let $\omega$ denote the EPR
state of $\alg{A}$.  Let $(\pi ,\hil{H},\Omega )$ denote the GNS
representation of $\alg{A}$ induced by $\omega$, and let
$\mathbf{B}(\hil{H})$ denote the algebra of bounded operators on
$\hil{H}$.  Then, $\alg{R}_{1}\equiv \pi (\alg{A}_{1})''$ and
$\alg{R}_{2}\equiv \pi (\alg{A}_{2})''$ are von Neumann subalgebras of
$\mathbf{B}(\hil{H})$ such that $\alg{R}_{1}\subseteq \alg{R}_{2}'$.
We first note an important fact about the relationship between the
vector $\Omega$ and the algebras $\alg{R}_{1}, \alg{R}_{2}$.

\begin{lemma} $\Omega$ is a separating trace vector for $\alg{R}_{1}$
  (respectively, for $\alg{R}_{2}$).  \end{lemma}

\begin{proof} We show first that $\Omega$ is cyclic for
  $\alg{R}_{1}$.  By the GNS construction, $\Omega$ is cyclic for $\pi
  (\alg{A})$.  Since $\alg{A}$ is generated by products of Weyl
  operators, it follows that the set
  \begin{equation} \hil{M}=\{ \pi (W(\mathbf{a})\otimes W(\mathbf{b}))
\Omega : \mathbf{a},\mathbf{b}\in \mathbb{R}^{2} \}
    .\end{equation} is a total set in $\hil{H}$.  Let \begin{equation}
\hil{N}=\{ \pi (W(\mathbf{a})\otimes I)\Omega :\mathbf{a} \in \mathbb{R}^{2} \}
.\end{equation}  Then it will suffice for our conclusion to show that
$\hil{N}$ is a total set in $\hil{H}$.  Let $\psi \in \hil{M}$.  
That is, $\psi = \pi(W(a,b)\otimes
W(c,d))\Omega$, for some quadruple $a,b,c,d$ of real numbers.  Let
$\phi=\pi (W(a+c,b-d)\otimes I)\Omega \in \hil{N}$.  Note that since
Weyl operators are unitary, $\norm{\psi}=\norm{\phi}=1$.  Now,
\begin{eqnarray} \lefteqn{ [W(a,b)\otimes W(c,d)]^{*}[W(a+c,b-d)\otimes
  I] } \qquad \qquad \qquad \qquad \qquad \nonumber \\
&=& e^{it}W(c,-d)\otimes W(-c,-d) ,\end{eqnarray} where $t=(ad+bc)/2$.  Thus,
\begin{eqnarray}
\langle \psi ,\phi \rangle &=& \omega
  ([W(a,b)\otimes W(c,d)]^{*}[W(a+c,b-d)\otimes I]) \\
  &=& e^{it} \omega (W(c,-d)\otimes W(-c,-d)) \: =\: e^{it} 
e^{ic\lambda }e^{-id\mu } .\end{eqnarray}  Hence,
  $|\langle \psi ,\phi \rangle |=1$; that is, $\phi$ is a scalar multiple
  of $\psi$.  Since $\hil{M}$ is a total set
in $\hil{H}$ and since $\psi$ was an arbitrary vector in $\hil{M}$, it
follows that $\hil{N}$ is a total set in $\hil{H}$.  Therefore,
$\Omega$ is cyclic for $\alg{R}_{1}$.  By symmetry, $\Omega$ is cyclic
for $\alg{R}_{2}$.  Since $\alg{R}_{1}\subseteq \alg{R}_{2}'$,
$\Omega$ is separating for $\alg{R}_{1}$ and $\alg{R}_{2}$.    
  
In order to show that $\Omega$ is a trace vector for $\alg{R}_{1}$,
let $\mathbf{a},\mathbf{b}\in \mathbb{R}^{2}$.  If
$\mathbf{a}=-\mathbf{b}$, then $W(\mathbf{a})W(\mathbf{b})\otimes
I=W(\mathbf{b})W(\mathbf{a})\otimes I$.  If $\mathbf{a}\neq
-\mathbf{b}$, then
\begin{equation} 
\omega (W(\mathbf{a})W(\mathbf{b})\otimes I) = \exp \{ i\sigma
    (\mathbf{a},\mathbf{b}) \} \, \omega
    (W(\mathbf{a}+\mathbf{b})\otimes I) =
    0 .\end{equation} 
Similarly, 
$\omega (W(\mathbf{b})W(\mathbf{a})\otimes I)=0$.  In
    either case, 
\begin{equation}  
\omega (W(\mathbf{a})W(\mathbf{b})\otimes I)=\omega
(W(\mathbf{b})W(\mathbf{a})\otimes I) .\end{equation} By taking linear
combinations of Weyl operators and norm limits, it follows that $\omega$ is a tracial
state of $\weyl \otimes I$.  By taking weak limits in the GNS
representation, it follows that $\Omega$ is a trace vector for
$\alg{R}_{1}$.  By symmetry, $\Omega$ is a trace vector for
$\alg{R}_{2}$.  \end{proof}
 
The previous lemma shows that $\alg{R}_{i}$ has a faithful numerical
trace, and is therefore a finite von Neumann
algebra~\cite[p.~504]{kr}.  Thus, $\pi |_{\alg{A}_{i}}$ is the unique
(up to quasi-equivalence) representation of $\weyl$ that generates a
finite von Neumann algebra, and $\alg{R}_{i}$ is a factor of type
II$_{1}$~\cite[Prop.~3.4]{slawny}.  It also follows that
$\alg{R}_{2}=\alg{R}_{1}'$.  Indeed, if the inclusion
$\alg{R}_{2}\subseteq \alg{R}_{1}'$ were proper, then it would follow
that $\alg{R}_{2}$ is of infinite type~\cite[Lemma~2]{kadison} (since
$\Omega$ is cyclic and separating for both $\alg{R}_{2}$ and
$\alg{R}_{1}'$).  Therefore, $\alg{R}_{2}=\alg{R}_{1}'$.

\begin{thm} The EPR state is maximally Bell correlated across 
  \mbox{$\weyl \otimes \weyl$}. \label{bell} \end{thm}

\begin{proof} Let $(\pi ,\hil{H},\Omega )$ be the GNS
  representation of $\alg{A}$ induced by $\omega$, and let
  $\alg{R}_{i}=\pi (\alg{A}_{i})''$.  By Lemma~\ref{cstar}, it will
  suffice to find a Bell operator $R$ for $\alg{R}_{1},\alg{R}_{2}$
  such that $\langle \Omega ,R\Omega \rangle=\sqrt{2}$.  Since
  $\alg{R}_{1}$ is type II, there is a projection $P\in \alg{R}_{1}$
  such that $P$ is equivalent to $I-P$~\cite[Lemma~6.5.6]{kr}.  That
  is, there is a partial isometry $V\in \alg{R}_{1}$ such that
  $VV^{*}=P$ and $V^{*}V=I-P$.  [Note that $V^{2}=(V^{*})^{2}=0$.]
  For each $\theta\in \mathbb{R}$, the operator $A(\theta )\equiv \exp
  (i\theta )V+\exp (-i\theta )V^{*}$ is self-adjoint and unitary.
  Moreover, for $\theta _{1},\theta _{2}\in \mathbb{R}$, we have
  \begin{equation} A(\theta _{1})A(\theta _{2})=\exp \{ i(\theta
    _{1}-\theta _{2})\} P +\exp \{ -i(\theta _{1}-\theta _{2})\} (I-P).
  \end{equation} Since $\Omega$ is a trace vector for $\alg{R}_{1}$, it
  follows that $\langle \Omega ,P\Omega \rangle =\langle \Omega
  ,(I-P)\Omega \rangle$.  Hence, $\langle \Omega ,P\Omega \rangle =
  1/2$, and
\begin{equation}
\langle \Omega ,A(\theta _{1})A(\theta _{2})\Omega \rangle = \cos
(\theta _{1}-\theta _{2}) .\end{equation}
Since $\Omega$ is also cyclic for $\alg{R}_{1}$, there is a $*$
anti-isomorphism $\gamma$ of $\alg{R}_{1}$ onto $\alg{R}_{1}'$ such
that $\gamma (A)\Omega =A\Omega$ for all $A\in \alg{R}_{1}$~\cite[Theorem
7.2.15]{kr}.  Defining self-adjoint unitaries $A_{i}\in \alg{R}_{1},
B_{i}\in \alg{R}_{1}'=\alg{R}_{2}\;(i=1,2)$ by $A_{1}=A(0), A_{2}=A(\pi
/2), B_{1}=\gamma (A(\pi /4)), B_{2}=\gamma (A(-\pi /4))$, one obtains
(cf.~\cite[Theorem 2.1]{cmp})
\begin{equation}
{\textstyle \frac{1}{2} }\langle \Omega
,(A_{1}(B_{1}+B_{2})+A_{2}(B_{1}-B_{2}))\Omega \rangle =2\cos (\pi /4)
=\sqrt{2} .\end{equation} \end{proof}

\begin{remark}  Since representations of $\alg{A}\equiv \weyl \otimes 
  \weyl$ are faithful, there is in each representation $(\pi
  ,\hil{H})$ of $\alg{A}$ a state (perhaps non-normal) corresponding
  to the EPR state.  That is, there is a state $\tau$ of $\pi
  (\alg{A})$ such that
\begin{equation}
\tau (\pi (A))=\omega (A) ,\qquad A\in \alg{A} .\end{equation}
In particular, this recipe may be applied to define an EPR state
$\tau$ in the Schr{\"o}dinger representation $\pi _{s}$
(cf.~\cite[Section 4.3]{fink}).  However, since all states in the folium of
$\pi _{s}$ are regular, and $\omega$ is non-regular, it follows that 
$\tau$ is non-normal.  Nonetheless, since $\pi _{s}(\weyl \otimes I)\subseteq
\mathbf{B}(\hil{H}_{1})\otimes I$ and $\pi _{s}(I\otimes \weyl)\subseteq
I\otimes \mathbf{B}(\hil{H}_{2})$, the singular state $\tau$ is maximally  
Bell correlated across $\mathbf{B}(\hil{H}_{1})\otimes
\mathbf{B}(\hil{H}_{2})$.  \end{remark}

\section{Characterizing EPR-type states}
According to a recent scheme of Arens and Varadarajan~\cite{vara}, a
state $\rho$ of a joint quantum system $S_{1}\times S_{2}$, with
Hilbert space $\hil{H}_{1}\otimes \hil{H}_{2}$, is said to be an
\emph{EPR-type state} for an observable $A\in
\mathbf{B}(\hil{H}_{1})\otimes I$ just in case there is some
observable $A'\in I\otimes \mathbf{B}(\hil{H}_{2})$ such that the
joint distribution of $A$ and $A'$ with respect to $\rho$ is
concentrated on the diagonal; that is, $\rho ((A-A')^{2})=0$.  In such
a case, we say that $A'$ is a \emph{double} of $A$ relative to the
state $\rho$.

As they themselves acknowledge, Arens and Varadarajan's
characterization is not sufficiently general to cover the original EPR
state.  In particular, they assume that $\rho$ is a normal state on a
tensor product of type I factors---which the original EPR state is
not---and it follows from this that any observable with a double
relative to $\rho$ must have a discrete spectrum~\cite[Theorem
4]{vara}.
 
However, the characterization of EPR-type states by Arens and
Varadarajan has been extended by Werner~\cite{werner} to the case
where the observables of the subsystems are given, respectively, by a
von Neumann algebra $\alg{R}$ and its commutant $\alg{R}'$.  Let
$D(\alg{R},\alg{R}',\rho )$ denote the set of elements in $\alg{R}$
for which a double exists in $\alg{R}'$ relative to $\rho$.  Let
$C_{\rho}(\alg{R})$ denote the centralizer of $\rho$ in $\alg{R}$.
That is, $C_{\rho}(\alg{R})$ consists of those elements $A\in \alg{R}$
such that $\rho (AB)=\rho(BA)$ for all $B\in \alg{R}$.  We then have
the following result:

\begin{thm}[\cite{werner}] Let $\alg{R}$ be a von Neumann algebra with
  cyclic and separating vector $\Omega$, and let $\rho$ be the state
  induced by $\Omega$.  Then \[ D(\alg{R},\alg{R}',\rho
  )=C_{\rho}(\alg{R}) .\] Moreover, the double $A'\in \alg{R}'$ of any
  $A\in C_{\rho}(\alg{R})$ is unique.
\label{werner} \end{thm}
It is possible, in general, for an observable $A\in C_{\rho}(\alg{R})$
to have a continuous spectrum (see, for example, the following
paragraph).  If, however, $\alg{R}$ is type I, then
$C_{\rho}(\alg{R})$ contains only discrete spectrum
observables~\cite{werner}, re-establishing the conclusion of Arens and
Varadarajan.

In the case of present interest, we let $\alg{R}=\pi (\weyl \otimes
I)''$, where $(\pi ,\hil{H},\Omega )$ is the GNS representation of
$\weyl \otimes \weyl$ induced by the EPR state.  As we saw previously,
it then follows that $\alg{R}'=\pi (I\otimes \weyl )''$.  Let $\rho$
be the state of $\mathbf{B}(\hil{H})$ induced by $\Omega$.  Since
$\Omega$ is cyclic and separating for $\alg{R}$, it follows from
Theorem~\ref{werner} that $D(\alg{R},\alg{R}',\rho
)=C_{\rho}(\alg{R})$.  Moreover, since $\Omega$ is a trace vector for
$\alg{R}$, it follows that $C_{\rho}(\alg{R})=\alg{R}$, and every
element $A\in \alg{R}$ has a unique double $A'\in \alg{R}'$.  [In
fact, $A'=\gamma (A^{*})$, where $\gamma$ is the $*$ anti-isomorphism
invoked in the proof of Theorem~\ref{bell}.]  Therefore, the original
EPR state does qualify as an EPR-type state in the sense of Werner.

Finally, it is now clear that the heuristic analogy between the
original EPR state and Bohm's singlet spin-$1/2$ state has more than a
superficial basis.  Indeed, in both cases the state $\rho$ is induced
by a vector that is cyclic and separating for the component observable
algebras $\alg{R},\alg{R}'$.  And in each case $\rho$ restricts to the
unique tracial state on the component algebras.  These two facts, in
turn, entail that $\rho$ is maximally Bell correlated across
$\alg{R},\alg{R}'$, and that for each observable $A\in \alg{R}$ there
is a unique observable in $A'\in \alg{R}'$ such that $A$ and $A'$ are
perfectly correlated in the state $\rho$.

\vspace{0.5em} \noindent {\bf Acknowledgment:} I am indebted to Rob
Clifton for insightful discussions and encouragement during this
project.
 
 \end{document}